\documentclass[11pt]{article}
\usepackage[margin=1in]{geometry}
\usepackage[T1]{fontenc}
\usepackage{lmodern}
\usepackage{microtype}
\usepackage{amsmath,amssymb}
\usepackage{graphicx}
\usepackage{booktabs}
\usepackage{tabularx}
\usepackage{longtable}
\usepackage{array}
\usepackage{multirow}
\usepackage{xcolor}
\usepackage{enumitem}
\usepackage{tikz}
\usetikzlibrary{arrows.meta,positioning,fit,backgrounds,shapes.geometric}
\usepackage[numbers,sort&compress]{natbib}
\usepackage[colorlinks=true,linkcolor=blue!50!black,citecolor=blue!50!black,urlcolor=blue!50!black]{hyperref}
\usepackage{caption}
\setlist{itemsep=2pt,topsep=4pt}
\newcommand{\skill}[1]{\texttt{#1}}

\title{\textbf{Mechanical Engineering Skills: An Open Library of Agent Skills\\ for Design, Analysis and Manufacturing}}

\author{
Achuth Chandrasekhar\textsuperscript{1} \quad Jesse Barkley\textsuperscript{1} \quad Peter Pak\textsuperscript{1} \quad Rumi Loghmani\textsuperscript{1} \\
 \quad Aakash Oza\textsuperscript{1} \quad Amir Barati Farimani\textsuperscript{1,2,3,*} \\[6pt]
\small \textsuperscript{1}Department of Mechanical Engineering, Carnegie Mellon University, Pittsburgh, PA 15213, USA \\
\small \textsuperscript{2}Department of Chemical Engineering, Carnegie Mellon University, Pittsburgh, PA 15213, USA \\
\small \textsuperscript{3}Machine Learning Department, Carnegie Mellon University, Pittsburgh, PA 15213, USA \\[2pt]
\small \textsuperscript{*}Corresponding author: \texttt{barati@cmu.edu}
}
\date{}

\begin{document}
\maketitle

\begin{abstract}
Mechanical engineers rely on dozens of specialized tools---CAD kernels, finite-element and CFD
solvers, flight simulators, process maps, thermodynamic databases and research models---each with
its own interface, installation and learning curve. We present \emph{Mechanical Engineering
Skills}, an open library of 33 agent skills for design, analysis and manufacturing that makes this
breadth available through a single conversation, by text or, in hosts that support it, by voice.
Each skill packages operating instructions, executable tools and verification procedures: the
language model understands the request, selects the skill and coordinates the workflow, while
deterministic tools do the computation and results are returned with the checks that support them.
For the default workflows, users do not need to learn each tool's interface, train a model, or run
a model on their own computer; skills load only when a request needs them and install with one
command. Nineteen skills adapt published or preprint research from our laboratory, spanning
text-to-CAD, linkage synthesis, CFD case setup, stress-field and lattice surrogates, equation
discovery, additive-manufacturing process planning and in-process monitoring; 14 cover core
engineering and teaching tasks. We describe the skill format, the design principles and a procedure
for translating research into usable workflows. Five case studies, each with an explicit evidence
scope and an independent rerun, show exact analytical agreement for lattice homogenization, rocket
apogee within 4\% of stored OpenRocket references, a reproduced equation fit that is nine times more
accurate than a cubic polynomial at an unseen temperature, and a porosity-monitoring pipeline
rehearsed end to end. A controlled comparison of agents with and without skills is left to future
work. The library is available at
\url{https://github.com/BaratiLab/mechanical-engineering-skills}.
\end{abstract}

\section{Introduction}
Mechanical engineering work is a chain of small, checkable steps: read a requirement, pick a
model, compute, compare against a hand estimate or a known case, and decide. General-purpose
language-model agents are now capable of driving such chains: they write CAD scripts, set up
simulations, and read sensor data. Our own work has shown this across the life cycle of a part,
from text-to-CAD~\cite{barkley2026cadsmith,badagabettu2024query2cad} and mechanism
synthesis~\cite{jadhav2026linkd} to LLM-in-the-loop structural design~\cite{jadhav2026designer},
autonomous CFD~\cite{oza2026cfdforge}, additive-manufacturing process
planning~\cite{pak2025additivellm,pak2026alloy,pak2026sls} and print
monitoring~\cite{jadhav2025llm3dprint}.

Two obstacles keep this research from becoming everyday practice. First, each method lives in its
own code base, with its own environment, data and conventions, and is rarely packaged for someone
who simply has an engineering question. Second, agents left to themselves make the kind of errors
engineers cannot accept: arithmetic slips, wrong units, a correlation used outside its range, or a
confident answer with no evidence behind it.

Agent Skills~\cite{anthropic2025skills,agentskills2025spec} offer a lightweight remedy. A skill is a
directory with a \texttt{SKILL.md} file---a short description that tells the agent when to use
it, followed by instructions---plus any scripts and reference files the instructions point to. The
agent sees only the one-line descriptions until a task matches, then loads the instructions, and
reads deeper files or runs scripts only when needed (\emph{progressive disclosure}). Skills are
portable across Claude Code, Claude.ai and other agents that adopted the format, and they are
installed with a single command. Numerical packages, external solvers and any model weights are installed separately (\S\ref{sec:access}).

\begin{figure}[t]
  \centering
  \includegraphics[width=\textwidth]{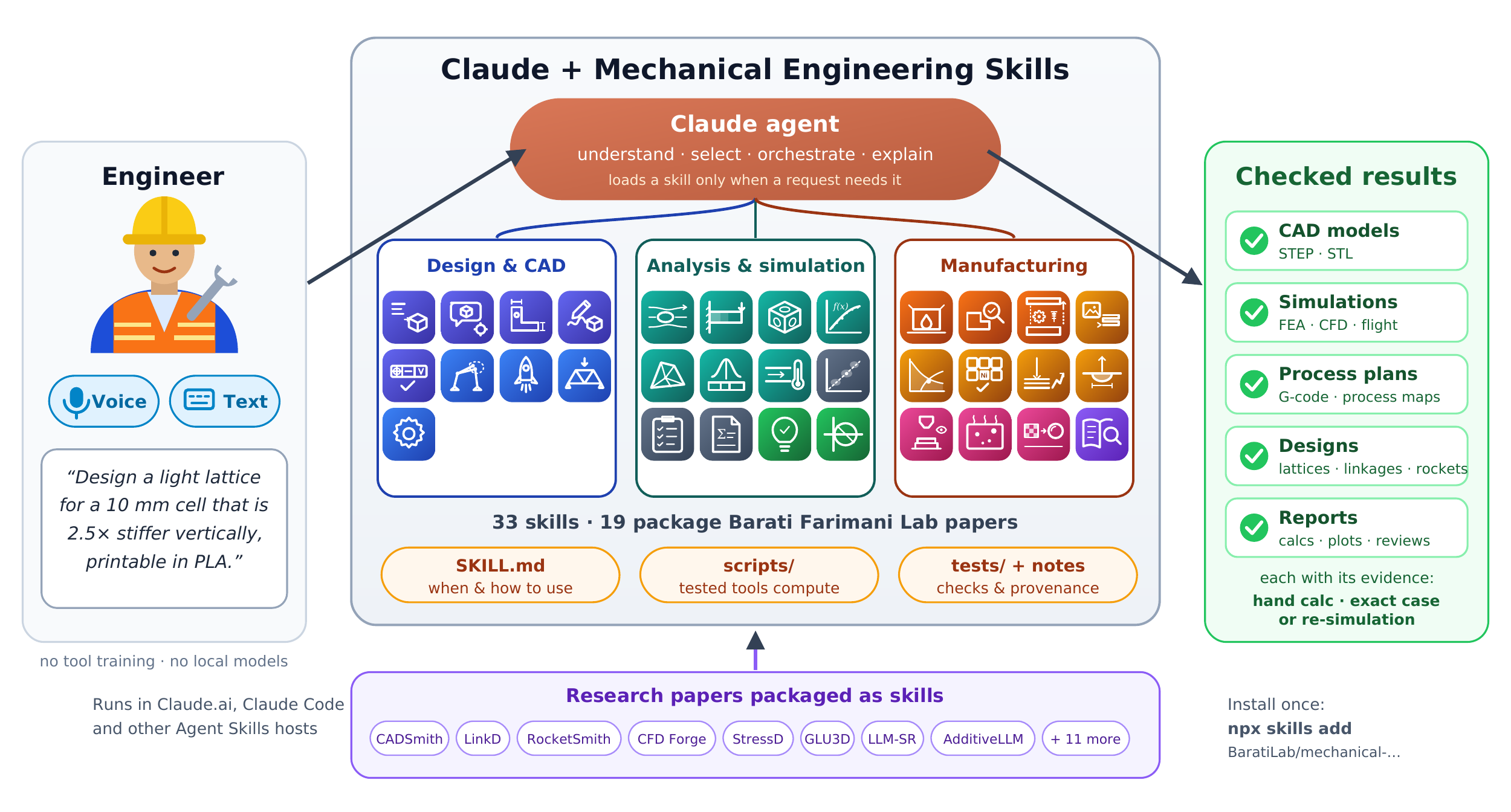}
  \caption{Mechanical Engineering Skills. An engineer states a goal by voice or text.
  The Claude agent understands the request, selects the relevant skills from a library of 33
  (design and CAD, analysis and simulation, manufacturing), and orchestrates them. Each skill
  bundles instructions (\texttt{SKILL.md}), tested scripts that do the computation, and checks.
  The results---CAD models, simulations, process plans, designs and reports---come back with the
  evidence that supports them. Nineteen skills package research papers from our laboratory. No
  per-tool training or locally hosted model is required for the default workflows; voice input and
  execution depend on the host and the installed tools.}
  \label{fig:concept}
\end{figure}

This paper presents \emph{Mechanical Engineering Skills}~\cite{mes2026repo}, a library of 33 such
skills (Fig.~\ref{fig:concept}, Table~\ref{tab:library}). Our contributions are:
\begin{enumerate}[leftmargin=*]
\item \textbf{A library} covering design, analysis and manufacturing, in which 19 skills package
      peer-reviewed or preprint research from our laboratory and 14 cover core engineering tasks
      (machine-element sizing, FEA with hand checks, tolerance stack-ups, material selection,
      design reviews and teaching aids).
\item \textbf{Design principles} for engineering skills: tools compute and the model orchestrates;
      consequential numbers are tied to explicit checks; changes to a source method are declared;
      and the skill states the boundary of its validity.
\item \textbf{Access without expertise in each tool}: the whole library is reachable by text or
      voice in one conversation with Claude, without learning tool interfaces, training models or
      running models locally, illustrated by twelve end-to-end tasks (\S\ref{sec:access}).
\item \textbf{Five case studies} drawn from our research---lattice inverse design, printed rockets,
      in-situ porosity monitoring, polymer process optimization and equation discovery---each
      with an explicitly stated evidence scope and confirmed by an independent rerun.
\end{enumerate}

\section{Motivation and prior work}
\label{sec:related}

Mechanical engineering is a demanding setting for language-model agents: its answers are numbers
with units and tolerances, they are produced by specialized tools, and a wrong one can break a part.
Three lines of work meet in this library.

Large language models acquire broad knowledge and
in-context reasoning from scale~\cite{brown2020gpt3,wei2022cot}, and become agents when they
interleave reasoning with actions and observations~\cite{yao2023react} or learn to call external
tools~\cite{schick2023toolformer}; surveys catalogue the resulting architectures~\cite{wang2024agentsurvey}.
Retrieval augments generation with documents the model was not trained on~\cite{lewis2020rag},
and the Model Context Protocol standardizes connections to external systems~\cite{anthropic2024mcp}.
Agent Skills~\cite{anthropic2025skills,agentskills2025spec} add a complementary layer: packaged
procedural knowledge and code that an agent loads only when relevant. In science, tool-using
agents have planned and executed chemistry experiments~\cite{boiko2023coscientist,bran2024chemcrow};
our library brings the same pattern to mechanical engineering, with an emphasis on checks.

Early studies showed that
general models can help across the design-to-manufacture pipeline but make errors that require
verification~\cite{makatura2023llmdesign}. Subsequent work generated CAD from text with execution
feedback~\cite{badagabettu2024query2cad,barkley2026cadsmith}, coupled models with finite-element
analysis for structural design~\cite{jadhav2026designer}, automated CFD setup under physics
constraints~\cite{oza2026cfdforge}, and applied agents to additive manufacturing, from literature
question answering~\cite{chandrasekhar2024amgpt} and defect prediction~\cite{pak2025additivellm}
to alloy evaluation~\cite{pak2026alloy}, process optimization~\cite{pak2026sls}, print monitoring
and control~\cite{jadhav2025llm3dprint} and complete design-to-flight workflows~\cite{pak2026rocketsmith}.
These systems are typically bespoke; this work packages them in a common, portable format.

Generative models, in particular denoising
diffusion~\cite{ho2020ddpm,nichol2021improved,song2021ddim,ho2022cfg}, now produce stress
fields~\cite{jadhav2023stressd}, lattice architectures~\cite{jadhav2024glu3d}, mechanisms~\cite{jadhav2026linkd},
toolpaths~\cite{wang2026image2gcode} and super-resolved process images~\cite{ogoke2025superres},
often with U-Net backbones~\cite{ronneberger2015unet}. Machine learning also predicts melt-pool
geometry~\cite{akbari2022meltpoolnet} and porosity~\cite{pak2024thermopore} in metal additive
manufacturing~\cite{debroy2018am,king2014keyhole}, and symbolic regression recovers interpretable
laws from data~\cite{schmidt2009distilling,udrescu2020aifeynman,shojaee2025llmsr}. The skills pair
such models with classical references---Rosenthal's moving-source solution~\cite{rosenthal1946},
Barrowman's stability method~\cite{barrowman1967}, open-source solvers~\cite{weller1998openfoam,niskanen2009openrocket},
computational thermodynamics~\cite{andersson2002thermocalc}, cellular-solid
scaling~\cite{gibson1997cellular,alketan2019tpms}, numerical homogenization~\cite{dong2019homo3d}
and Bayesian optimization~\cite{jones1998ego}---so that learned predictions can be checked against
physics wherever the assumptions are comparable.

\section{Mechanical Engineering Skills}
\label{sec:anatomy}
\label{sec:library}
\label{sec:principles}

\begin{figure}[t]
\centering
\begin{tikzpicture}[
  font=\small,
  box/.style={draw, rounded corners=3pt, align=left, inner sep=5pt, fill=#1},
  box/.default=white,
  lbl/.style={font=\footnotesize\bfseries, text=black!70},
  arr/.style={-{Stealth[length=2.2mm]}, thick, black!60}]
  \node[box=blue!6, minimum width=3.3cm, align=center] (user) {Engineer's request\\ \emph{``a lattice with}\\ \emph{E$_3$ = 2.5 E$_1$, light''}};
  \node[box=blue!12, below=6mm of user, minimum width=3.3cm, align=center] (agent) {\textbf{Agent (Claude)}\\ plans, calls tools,\\ reads results, reports};
  \node[box=green!10, below=6mm of agent, minimum width=3.3cm, align=center] (out) {Checked deliverables\\ STEP/STL, G-code, plots,\\ report + evidence};
  \draw[arr] (user) -- (agent); \draw[arr] (agent) -- (out);
  \node[box=orange!8, right=22mm of agent, yshift=10mm, text width=6.6cm, align=left] (skillmd)
     {\texttt{SKILL.md}\\[1pt]\footnotesize \textbf{description}: when to use (always loaded)\\
      \textbf{workflow}: steps, questions to ask, defaults\\ \textbf{judgement calls}: validity limits};
  \node[box=orange!4, below=2mm of skillmd, text width=6.6cm] (scripts)
     {\texttt{scripts/}\ \footnotesize deterministic tools that compute (FEM, solvers, fits, CAD, G-code)};
  \node[box=orange!4, below=2mm of scripts, text width=6.6cm] (refs)
     {\texttt{references/}\ \footnotesize paper notes, data provenance, deviations};
  \node[box=orange!4, below=2mm of refs, text width=6.6cm] (tests)
     {\texttt{tests/}\ \footnotesize exact cases, paper numbers, regression checks};
  \begin{scope}[on background layer]
    \node[draw=orange!60, dashed, rounded corners=5pt, inner sep=6pt, fit=(skillmd)(tests),
          label={[lbl]above:skill folder (one per capability)}] (folder) {};
  \end{scope}
  \draw[arr] (agent.east) -- node[above, font=\footnotesize, align=center] {loads when\\relevant} (folder.west|-agent.east);
  \draw[arr, dashed] (scripts.west) to[bend left=18] node[below, font=\footnotesize, pos=0.4] {runs} (agent.south east);
  \node[box=gray!8, right=5mm of folder, text width=2.2cm, align=center] (tut) {Tutorial + use cases\\ \footnotesize (project website)};
  \draw[arr, gray] (folder.east) -- (tut.west);
\end{tikzpicture}
\caption{Anatomy of an engineering skill and how an agent uses it. Only the one-line description
is always in the agent's context; instructions, scripts and references are loaded on demand. Tests
and a public tutorial accompany every research skill.}
\label{fig:anatomy}
\end{figure}
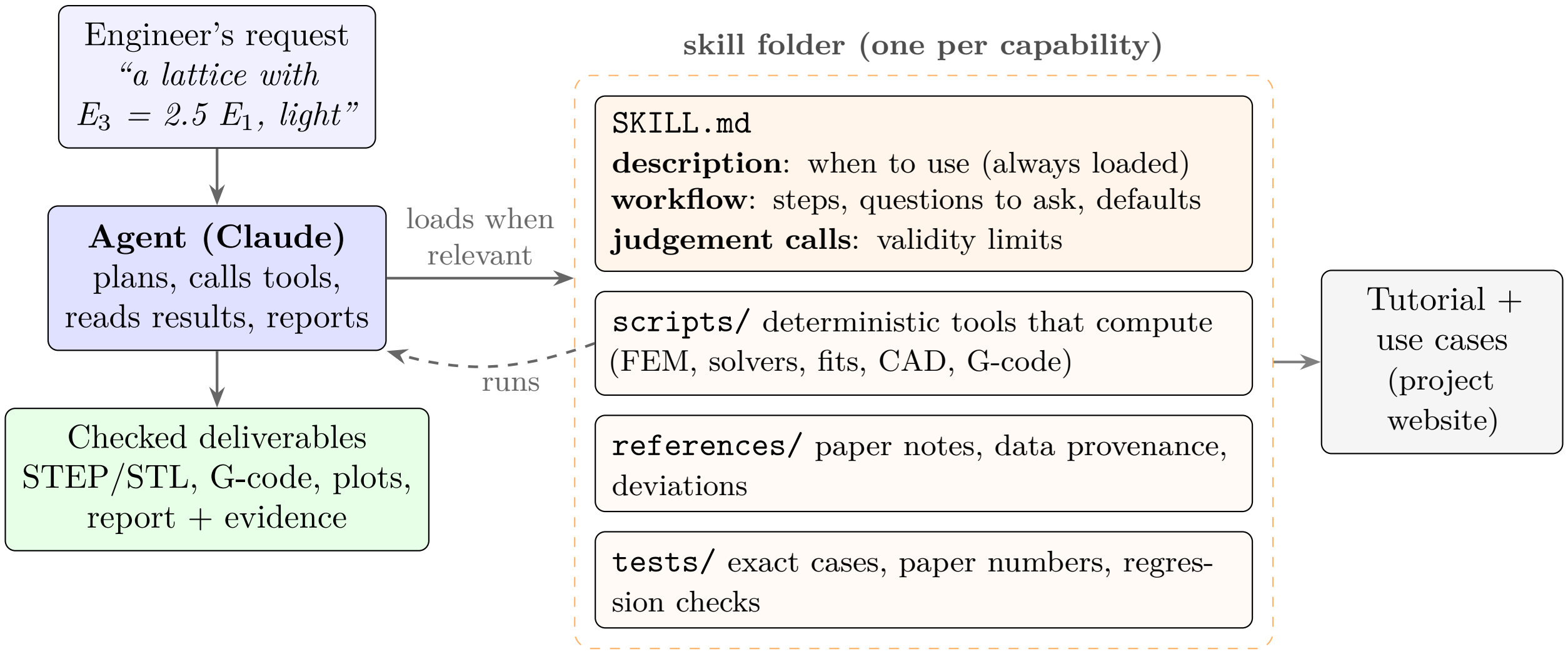
Figure~\ref{fig:anatomy} shows the structure we use for every skill. The \texttt{SKILL.md}
front matter carries a description written for the agent's routing decision: what the skill does,
what it is based on, and the situations that should trigger it. The body is a short
workflow---typically four to six numbered steps---with the commands to run and, crucially, the
\emph{judgement calls} an experienced engineer would make: which inputs to ask for, what to report
alongside the headline number, and where the method stops being valid. Long material (paper
notes, provenance, interpretation choices) sits in \texttt{references/} so that it costs no
context unless needed.

The \texttt{scripts/} directory holds the computation. Scripts are ordinary command-line Python
programs with JSON output so that the agent can parse results reliably. Examples include a
periodic voxel homogenizer for lattices, a plane-stress Q4 finite-element solver, a Barrowman
stability and one-dimensional flight simulator for rockets, a Rosenthal melt-pool solver, an
SQLite build log with Gaussian-process suggestions for selective laser sintering, and validators
for OpenFOAM cases. Each research skill also ships a test suite (\texttt{tests/}) and a tutorial
on the project website with executable examples and a statement of where their data come from.

Five principles shape every skill. The first is that the model orchestrates while tools compute:
instructions require each consequential number to come from an executable calculation with declared
inputs and units, never from the model's own arithmetic---the conclusion RocketSmith~\cite{pak2026rocketsmith}
reached after flying four agent-designed rockets, which we apply library-wide. The model remains
responsible for what it does well: eliciting requirements, choosing a method, chaining tools and
explaining results. The second is that every result travels with its check. A finite-element run is
returned with a hand calculation, a lattice homogenization with the exact laminate solution, a
surrogate's best candidate with a re-simulation, a CAD edit with a kernel validity and volume check.
Where a published method accepted designs with a learned scorer, the skill accepts on the physics
model and keeps the learned model for guidance; if a check fails or cannot run, the result is reported
as unverified together with its inputs, assumptions and software versions, and agreement between
tools that share assumptions is treated as consistency rather than physical validation.

Research skills follow two further rules. They keep reproduction and adaptation apart, recording
which parts retain the source protocol, which use a replacement model and which have only been
exercised on synthetic data, with the reason for every changed default or acceptance criterion; a
working implementation is not, by itself, a reproduction of a paper's headline accuracy. And they state
their boundaries where the agent will repeat them: elastic homogenization is not crash performance, a
flight simulation is a design check rather than a launch approval, and a defect classifier trained on
one machine family is not a process qualification. Finally, skills ask for the inputs that matter and
then act. Instructions name the few inputs that change the engineering decision---printed modulus,
load cases, minimum wall thickness, certification level---and give defaults for the rest, so the agent
asks a small set of grouped questions instead of many generic ones and asks again only when a new
ambiguity would change the result.

The library itself is summarized in Table~\ref{tab:library}, which lists all 33 skills, grouped into design and CAD, analysis and simulation,
manufacturing, and teaching. Each row gives the skill's icon and name and, in plain language, what
a user can ask for. Nineteen skills package research from our laboratory; the remaining 14 cover
core engineering tasks that every practitioner needs, from machine-element sizing and FEA with hand
checks to tolerance stack-ups, material selection~\cite{ashby2011} and design reviews, plus two
skills for teaching.

{\small
\begin{longtable}{@{}>{\centering\arraybackslash}m{1.25cm} >{\raggedright\arraybackslash}m{4.5cm} >{\raggedright\arraybackslash}m{8.3cm} >{\centering\arraybackslash}m{1.1cm}@{}}
\caption{The skill library. Every skill is installed with one command and used by asking for the
result in plain language; the agent selects the skill from its description. Execution additionally requires
the skill's dependencies. A citation marks the 19 skills that package a Barati Farimani Lab paper.}\label{tab:library}\\
\toprule
 & \textbf{Skill} & \textbf{What you can ask for} & \textbf{Paper} \\
\midrule
\endfirsthead
\toprule
 & \textbf{Skill} & \textbf{What you can ask for} & \textbf{Paper} \\
\midrule
\endhead
\bottomrule
\endfoot
\multicolumn{4}{@{}l}{\rule{0pt}{14pt}\textbf{\large Design and CAD}}\\[3pt]
\includegraphics[width=1.0cm]{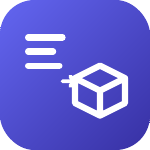} & \texttt{cadsmith-text-to-cad} & Describe a part in words; get a CadQuery model checked by exact geometry measurements and a three-view review, exported as STEP/STL. & \cite{barkley2026cadsmith} \\[1pt]
\includegraphics[width=1.0cm]{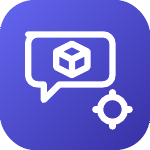} & \texttt{query2cad-freecad} & Plain-language request to a FreeCAD macro, with automatic error repair and a visual self-check. & \cite{badagabettu2024query2cad} \\[1pt]
\includegraphics[width=1.0cm]{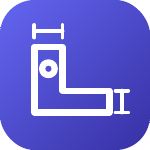} & \texttt{parametric-part-cad} & Part specification in plain English to a parametric CadQuery script, STEP/STL and a preview. & -- \\[2pt]
\includegraphics[width=1.0cm]{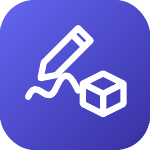} & \texttt{sketch-to-cad} & Hand sketch, photo or video frames to a dimensioned CAD model. & -- \\[2pt]
\includegraphics[width=1.0cm]{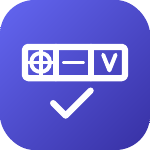} & \texttt{drawing-gdt-review} & Audit an engineering drawing for missing dimensions and GD\&T errors. & -- \\[2pt]
\includegraphics[width=1.0cm]{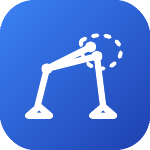} & \texttt{linkage-synthesis} & Target motion path to a four- or six-bar linkage, built joint by joint and checked kinematically. & \cite{jadhav2026linkd} \\[1pt]
\includegraphics[width=1.0cm]{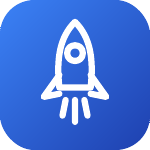} & \texttt{rocketsmith} & Hobby high-power rocket: stability, fin sizing, flight prediction, drag calibration and printable parts. & \cite{pak2026rocketsmith} \\[1pt]
\includegraphics[width=1.0cm]{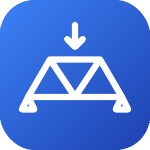} & \texttt{llm-mechanical-designer} & Truss design loop: the model proposes, exact FEM evaluates, the best designs are ranked. & \cite{jadhav2026designer} \\[1pt]
\includegraphics[width=1.0cm]{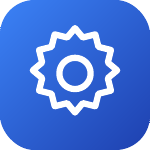} & \texttt{machine-element-sizing} & Bolts, shafts, bearings, springs, gears and welds, with the work shown and safety factors. & -- \\[2pt]
\multicolumn{4}{@{}l}{\rule{0pt}{14pt}\textbf{\large Analysis and simulation}}\\[3pt]
\includegraphics[width=1.0cm]{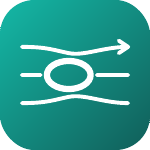} & \texttt{cfd-simulation-setup} & Flow problem to an OpenFOAM case, run and diagnosis; results accepted only against physics checks. & \cite{oza2026cfdforge} \\[1pt]
\includegraphics[width=1.0cm]{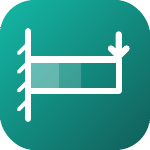} & \texttt{stress-field-surrogate} & Geometry, supports and loads to a full stress field in milliseconds, with its own FEM data generator. & \cite{jadhav2023stressd} \\[1pt]
\includegraphics[width=1.0cm]{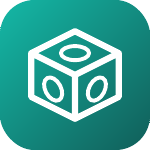} & \texttt{lattice-inverse-design} & Target stiffness to a printable TPMS or hybrid lattice cell with its full stiffness matrix and a tiled STL. & \cite{jadhav2024glu3d} \\[1pt]
\includegraphics[width=1.0cm]{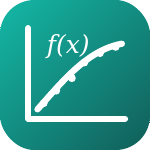} & \texttt{equation-discovery} & Test data to an interpretable equation, scored on a validation split and a final held-out test. & \cite{shojaee2025llmsr} \\[1pt]
\includegraphics[width=1.0cm]{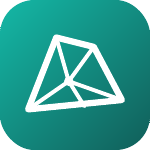} & \texttt{quick-fea} & Linear static and modal FEA with a required hand-calculation check. & -- \\[2pt]
\includegraphics[width=1.0cm]{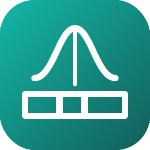} & \texttt{tolerance-stackup} & Worst-case, RSS and Monte Carlo tolerance stack-ups with contributor charts. & -- \\[2pt]
\includegraphics[width=1.0cm]{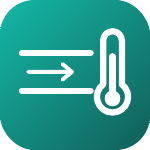} & \texttt{thermal-fluids-calc} & Pipe pressure drop, heat exchangers and fins with real fluid properties. & -- \\[2pt]
\includegraphics[width=1.0cm]{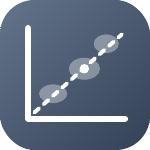} & \texttt{material-selection} & Performance indices to a ranked material shortlist with Ashby charts. & -- \\[2pt]
\includegraphics[width=1.0cm]{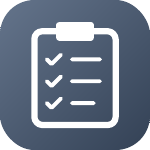} & \texttt{design-review-package} & Requirements, FMEA, bill-of-materials cost roll-up and review checklist. & -- \\[2pt]
\includegraphics[width=1.0cm]{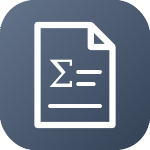} & \texttt{engineering-calc-report} & Turn calculations into a calculation report ready for sign-off. & -- \\[2pt]
\multicolumn{4}{@{}l}{\rule{0pt}{14pt}\textbf{\large Manufacturing}}\\[3pt]
\includegraphics[width=1.0cm]{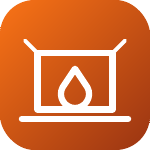} & \texttt{agentscad-fdm-dfam} & Make a STEP part printable by FDM: reorientation, teardrop holes and chamfers, with verified geometry. & \cite{george2026agentscad} \\[1pt]
\includegraphics[width=1.0cm]{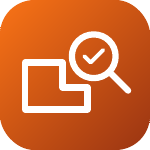} & \texttt{dfm-review} & Design-for-manufacture review for machining, sheet metal, molding and additive manufacturing. & -- \\[2pt]
\includegraphics[width=1.0cm]{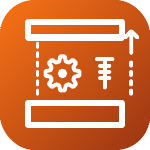} & \texttt{mechanical-teardown} & Product teardown: costed BOM, DFA index, should-cost, repairability and FMEA in an HTML report. & -- \\[2pt]
\includegraphics[width=1.0cm]{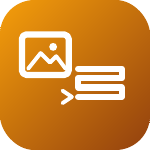} & \texttt{image2gcode} & Photo or sketch to printer-ready G-code for 2D and 2.5D parts. & \cite{wang2026image2gcode} \\[1pt]
\includegraphics[width=1.0cm]{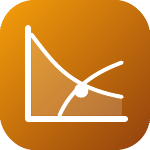} & \texttt{lpbf-defect-prediction} & Laser power and speed to keyholing, lack of fusion or balling, with process maps. & \cite{pak2025additivellm} \\[1pt]
\includegraphics[width=1.0cm]{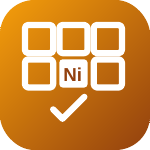} & \texttt{am-alloy-screening} & Known or new alloy to properties, a lack-of-fusion process map and a printability ranking. & \cite{pak2026alloy} \\[1pt]
\includegraphics[width=1.0cm]{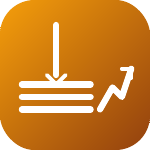} & \texttt{sls-process-optimization} & Polymer SLS test results to the next batch's settings, batch by batch. & \cite{pak2026sls} \\[1pt]
\includegraphics[width=1.0cm]{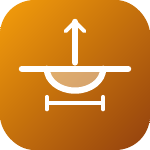} & \texttt{meltpool-geometry} & Melt-pool depth, width, length and mode from laser power, speed and alloy. & \cite{akbari2022meltpoolnet} \\[1pt]
\includegraphics[width=1.0cm]{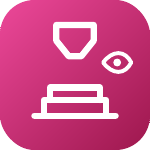} & \texttt{llm-3d-print-monitor} & Camera inspection of every layer and bounded in-print corrections on Klipper printers. & \cite{jadhav2025llm3dprint} \\[1pt]
\includegraphics[width=1.0cm]{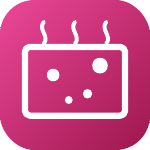} & \texttt{thermal-porosity} & Thermal images recorded during printing to pore count and location for each layer. & \cite{pak2024thermopore} \\[1pt]
\includegraphics[width=1.0cm]{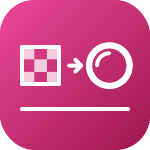} & \texttt{lpbf-superres} & Low-cost webcam images to high-resolution layer images and edge roughness. & \cite{ogoke2025superres} \\[1pt]
\includegraphics[width=1.0cm]{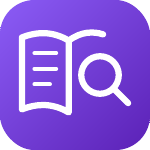} & \texttt{am-literature-rag} & Page-cited answers from your own additive-manufacturing paper library. & \cite{chandrasekhar2024amgpt} \\[1pt]
\multicolumn{4}{@{}l}{\rule{0pt}{14pt}\textbf{\large Teaching}}\\[3pt]
\includegraphics[width=1.0cm]{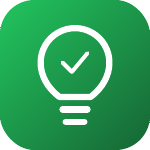} & \texttt{worked-problem-generator} & Textbook-style problems with code-verified solutions. & -- \\[2pt]
\includegraphics[width=1.0cm]{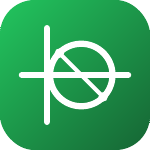} & \texttt{mechanics-visuals} & Free-body diagrams, Mohr's circle, beam diagrams and S--N curves. & -- \\[2pt]
\end{longtable}
}

\section{Engineering by voice or text with Claude}
\label{sec:access}

The practical value of a skill library is that it removes the cost of \emph{reaching} a capability.
Without it, using a research method or a specialized analysis means finding the code, installing its
environment, learning its inputs and conventions, often downloading or training a model and providing
a GPU, and then interpreting the outputs correctly. With skills, the engineer states the goal to
Claude---typed in Claude.ai or Claude Code, or spoken through voice mode in the Claude apps---and
Claude does the rest: it matches the request to a skill's description, loads that skill's
instructions, asks for the few inputs that matter, runs the scripts and returns the result with its
check.

Several properties make this practical. All 33 skills are installed with one command and are
available in the same conversation, and because only each skill's one-line description stays in the
model's context, on the order of a hundred tokens per skill, breadth costs almost nothing until a
skill is used. Skills compose naturally, so a lattice designed with \skill{lattice-inverse-design}
can be checked for overhangs with \skill{agentscad-fdm-dfam} and sliced without the engineer
switching tools. No training on each tool is needed, because the instructions encode what an expert
user of the underlying method knows---which inputs matter, which defaults are safe, which command to
run and how to read its output---so the engineer never has to learn the options of a homogenizer,
the input format of a flight simulator or the conventions of a thermodynamic database. The interface
is spoken language as much as text: an engineer at a printer or on the shop floor can ask,
hands-free, whether a parameter set risks lack of fusion or what fin span a rocket needs, and hear
the answer with its supporting check. Nor is there any model to download or run locally. The
language model runs in the cloud and the skills are text and lightweight scripts; where a source
paper relied on a large learned model, the default path uses physics models, solvers or small fitted
models that run in seconds on a CPU, such as search with exact homogenization instead of a 3D
diffusion model, or a Rosenthal solution calibrated on a few tracks instead of a trained network. In
Claude.ai the scripts run in Claude's own sandboxed code-execution environment, so nothing is
installed on the engineer's computer; in Claude Code they run in the engineer's project. Training the
larger research models remains possible, but it is never required to get an answer.

Some prerequisites remain. The installation command copies skill folders but does not install or
license every underlying tool: numerical scripts need Python packages such as NumPy and SciPy, CAD
workflows can need CadQuery or FreeCAD, CFD solves need OpenFOAM, thermodynamic calculations can need
a licensed database, and live monitoring needs supported printer hardware. Where a tool is missing,
skills fall back to validators, analytical alternatives or rehearsal data and say so. Voice reaches
the full workflow where the host also gives the agent access to the installed skills and to code
execution.

Table~\ref{tab:usecases} follows twelve tasks from the engineer's spoken or typed request, through
what Claude does with the skills, to the result it hands back, and Fig.~\ref{fig:dialogue} shows one
exchange in full.

{\small
\renewcommand{\arraystretch}{1.2}
\begin{longtable}{@{}>{\raggedright\arraybackslash}p{4.0cm} >{\raggedright\arraybackslash}p{6.7cm} >{\raggedright\arraybackslash}p{4.6cm}@{}}
\caption{Twelve engineering tasks requested of Claude by voice or text: the input, what Claude does
with the skills, and the final result returned to the engineer. The table illustrates intended use;
a controlled evaluation is future work (\S\ref{sec:limits}).}\label{tab:usecases}\\
\toprule
\textbf{Input (voice or text)} & \textbf{What Claude does} & \textbf{Final result} \\
\midrule
\endfirsthead
\toprule
\textbf{Input (voice or text)} & \textbf{What Claude does} & \textbf{Final result} \\
\midrule
\endhead
\bottomrule
\endfoot
{\scriptsize\textsc{voice}}\newline\emph{``Size the output shaft and bearings for a 2~kW gearbox at 300~rpm, steel, safety factor two.''} & \includegraphics[height=0.62cm]{figures/icons/machine-element-sizing.pdf}\hspace{3pt}\newline{\scriptsize \texttt{machine-element-sizing}}\newline Computes the torque (63.7~N\,m), sizes the shaft for combined bending and torsion fatigue, selects bearings for the required life and shows each step. & Shaft diameter and keyway, bearing designation and rated life, fatigue safety factors, and a calculation sheet. \\
\addlinespace[7pt]
{\scriptsize\textsc{text}}\newline\emph{``Make an L-bracket for an M6 bolt pattern on 40~mm centres, 3~mm wall, printable without supports.''} & \includegraphics[height=0.62cm]{figures/icons/cadsmith-text-to-cad.pdf}\hspace{3pt}\includegraphics[height=0.62cm]{figures/icons/agentscad-fdm-dfam.pdf}\hspace{3pt}\newline{\scriptsize \texttt{cadsmith-text-to-cad}, \texttt{agentscad-fdm-dfam}}\newline Writes a CadQuery model, measures it with the geometry kernel and a three-view check, then reorients it and adds teardrop holes and chamfers where overhangs need them. & Validated STEP and STL files, preview images and an overhang report. \\
\addlinespace[7pt]
{\scriptsize\textsc{voice}}\newline\emph{``Will this aluminium bracket deflect more than half a millimetre under a 500~N tip load? (STEP attached)''} & \includegraphics[height=0.62cm]{figures/icons/quick-fea.pdf}\hspace{3pt}\newline{\scriptsize \texttt{quick-fea}}\newline Meshes the part, runs a linear static and a modal analysis, and compares the deflection with a beam hand calculation before answering. & A yes/no answer with deflection, peak stress, safety factor, first modes and the hand-check agreement. \\
\addlinespace[7pt]
{\scriptsize\textsc{text}}\newline\emph{``Set up air flow over this heat-sink fin array at 2~m/s and tell me whether the result is converged.''} & \includegraphics[height=0.62cm]{figures/icons/cfd-simulation-setup.pdf}\hspace{3pt}\includegraphics[height=0.62cm]{figures/icons/thermal-fluids-calc.pdf}\hspace{3pt}\newline{\scriptsize \texttt{cfd-simulation-setup}, \texttt{thermal-fluids-calc}}\newline Builds and runs an OpenFOAM case, checks residuals, conservation and mesh sensitivity against registered criteria, and estimates the heat-transfer coefficient from correlations as a cross-check. & The case files, an accept-or-rerun decision with reasons, and pressure drop and heat transfer compared with the correlation. \\
\addlinespace[7pt]
{\scriptsize\textsc{voice}}\newline\emph{``I need a lattice infill for a 10~mm cell, about 2.5 times stiffer vertically than sideways, as light as possible, in PLA.''} & \includegraphics[height=0.62cm]{figures/icons/lattice-inverse-design.pdf}\hspace{3pt}\newline{\scriptsize \texttt{lattice-inverse-design}}\newline Asks for the absolute stiffness and minimum wall, searches homogenized TPMS and hybrid cells, refines thickness and re-checks the best cell at higher resolution. & The cell, its full stiffness matrix and weakest-direction modulus, and a tiled watertight STL (Fig.~\ref{fig:dialogue}). \\
\addlinespace[7pt]
{\scriptsize\textsc{voice}}\newline\emph{``I'm building a 98~mm rocket for an H motor. Will it be stable, and how high will it go?''} & \includegraphics[height=0.62cm]{figures/icons/rocketsmith.pdf}\hspace{3pt}\newline{\scriptsize \texttt{rocketsmith}}\newline Computes the centre of pressure by Barrowman's method, sizes the fins to the target stability, simulates the flight and checks rail-exit speed, thrust-to-weight and certification level. & Stability margin, predicted apogee and peak speed, a pass/fail checklist, and printable nose cone and fin can. \\
\addlinespace[7pt]
{\scriptsize\textsc{text}}\newline\emph{``Will 280~W at 900~mm/s keyhole on Ti-6Al-4V?''} & \includegraphics[height=0.62cm]{figures/icons/lpbf-defect-prediction.pdf}\hspace{3pt}\includegraphics[height=0.62cm]{figures/icons/meltpool-geometry.pdf}\hspace{3pt}\newline{\scriptsize \texttt{lpbf-defect-prediction}, \texttt{meltpool-geometry}}\newline Classifies the defect regime for the parameter set and estimates melt-pool depth and width from a calibrated moving-source model. & The predicted regime, melt-pool dimensions and a process map that marks a candidate operating window. \\
\addlinespace[7pt]
{\scriptsize\textsc{voice}}\newline\emph{``Layer 140 looks stringy on the camera---what is wrong and what should I change?''} & \includegraphics[height=0.62cm]{figures/icons/llm-3d-print-monitor.pdf}\hspace{3pt}\newline{\scriptsize \texttt{llm-3d-print-monitor}}\newline Reads the layer images and the printer state, diagnoses the likely cause and proposes G-code changes within safety bounds, waiting for approval before applying them. & A diagnosis, the approved correction applied to the running print, and a log of every change. \\
\addlinespace[7pt]
{\scriptsize\textsc{text}}\newline\emph{``Which layers of yesterday's build should go to CT first? (pyrometry and CT folders attached)''} & \includegraphics[height=0.62cm]{figures/icons/thermal-porosity.pdf}\hspace{3pt}\newline{\scriptsize \texttt{thermal-porosity}}\newline Filters and aligns the thermal frames, builds per-layer temperature and dwell maps, and runs the pore-count and localization models. & Pore counts and maps for every layer, ranked for inspection, with an alignment check image. \\
\addlinespace[7pt]
{\scriptsize\textsc{text}}\newline\emph{``Here are our last six SLS batches and test results. What should batch seven be?''} & \includegraphics[height=0.62cm]{figures/icons/sls-process-optimization.pdf}\hspace{3pt}\newline{\scriptsize \texttt{sls-process-optimization}}\newline Reduces the ASTM tensile and flexural data, updates the build log and proposes bounded next settings with the expected effect of each lever. & Fill energy, overlap and surface temperature for the next batch, with the rationale and a warning list from past builds. \\
\addlinespace[7pt]
{\scriptsize\textsc{text}}\newline\emph{``Find a law for this stress--strain data that also holds at a temperature I have not tested.''} & \includegraphics[height=0.62cm]{figures/icons/equation-discovery.pdf}\hspace{3pt}\newline{\scriptsize \texttt{equation-discovery}}\newline Proposes equation forms, fits their constants, ranks them on a validation split and reports the final held-out test once, for the frozen equation. & An interpretable equation with fitted constants, the physical meaning of each term and its errors. \\
\addlinespace[7pt]
{\scriptsize\textsc{voice}}\newline\emph{``Give my students three Mohr's-circle problems with worked solutions and figures.''} & \includegraphics[height=0.62cm]{figures/icons/worked-problem-generator.pdf}\hspace{3pt}\includegraphics[height=0.62cm]{figures/icons/mechanics-visuals.pdf}\hspace{3pt}\newline{\scriptsize \texttt{worked-problem-generator}, \texttt{mechanics-visuals}}\newline Writes the problems, solves them in code to verify every number, and draws the circles and stress elements. & Problem sheet, step-by-step solutions and figures, ready to hand out. \\
\end{longtable}
}

\begin{figure}[t]
\centering
\fbox{\begin{minipage}{0.94\textwidth}\small
\textbf{Engineer (voice):} \emph{``I need a lattice infill for a 10 mm cell that is about two and a
half times stiffer vertically than sideways, as light as possible, printed in PLA.''}\\[4pt]
\textbf{Agent:} loads \skill{lattice-inverse-design}; asks for the minimum printable wall and the
absolute stiffness needed. The engineer confirms $E_1=E_2\approx0.06\,E_s$ and $E_3\approx0.15\,E_s$
with printed $E_s=3.5$~GPa; the agent searches homogenized cells, refines thickness and re-checks the
best cell at higher resolution.\\[4pt]
\textbf{Agent (reply):} \emph{``A gyroid--diamond hybrid with two periods sideways gives
$E_3/E_1 = 2.6$ at 26\% density (about 0.2 and 0.5~GPa in PLA). Its weakest direction is still
$0.057\,E_s$, so it does not go soft off-axis. The tiled $3\times3\times3$ STL is attached.''}
\end{minipage}}
\caption{An illustrative exchange based on the lattice case study (\S\ref{sec:cases}). The engineer
names the goal; the skill supplies the procedure, the computation and the check. A stiffness ratio
alone does not fix a design, so the agent asks for absolute stiffness and the printable wall; the
search returns the best cell found, not a proven global minimum weight.}
\label{fig:dialogue}
\end{figure}
\section{Case studies}
\label{sec:cases}

Each case states what kind of evidence it rests on: an exact analytical check, a comparison with
published or stored reference values, an adaptation of the source method, or a rehearsal on
synthetic data. All numbers were recomputed in an independent rerun (\S\ref{sec:rerun}).

\subsection{Lattice inverse design (GLU3D)}
GLU3D~\cite{jadhav2024glu3d} generates lattice unit cells from a target stiffness with a 3D
conditional diffusion model trained on homogenized TPMS and hybrid-TPMS cells. The skill first
builds the physics: implicit cells, printability filters (one periodic solid that reaches every
face, no sealed cavities) and an energy-based periodic voxel homogenizer~\cite{dong2019homo3d} that
returns the full $6\times6$ stiffness. It uses periodic eight-node hexahedral elements, an
isotropic solid with $E_s=1$ and $\nu=0.3$ unless specified, Voigt order $(xx,yy,zz,yz,xz,xy)$ with
engineering shear strain, and omits void voxels. Directional moduli are evaluated from the
compliance tensor on 600 sphere directions, so the reported minimum is a sampled estimate.

\begin{figure}[t]
\centering
\begin{minipage}{0.56\textwidth}\centering
\includegraphics[width=\textwidth]{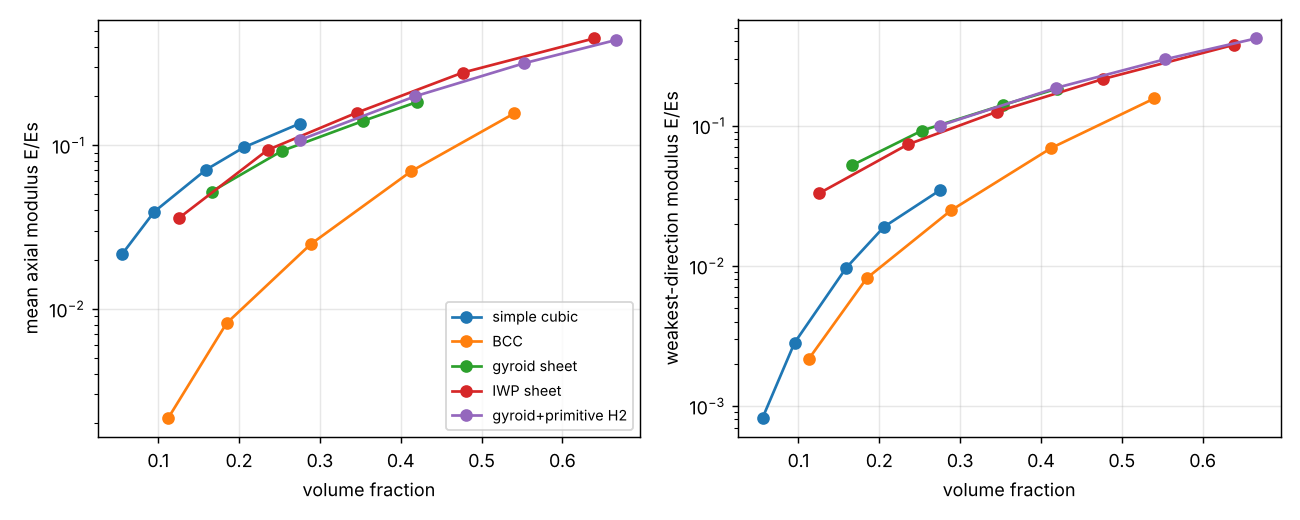}\\ \small (a)
\end{minipage}\hfill
\begin{minipage}{0.42\textwidth}\centering
\includegraphics[width=\textwidth]{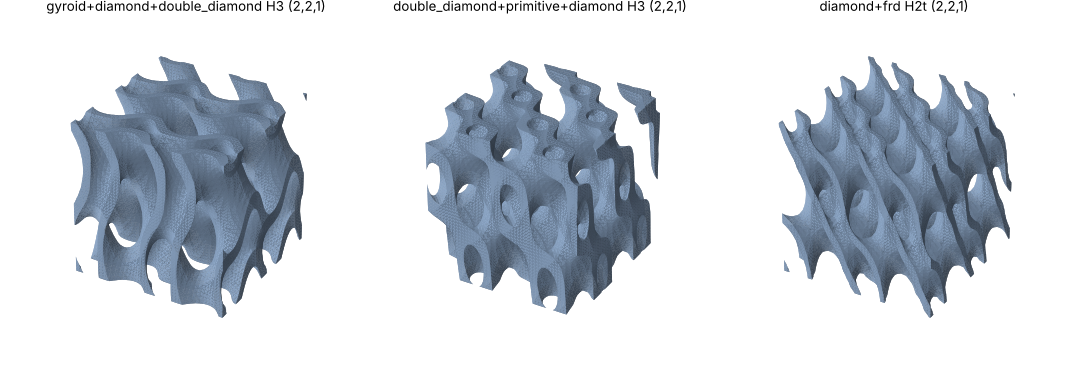}\\ \small (b)
\end{minipage}
\caption{Lattice inverse design. (a) Mean axial and weakest-direction modulus against volume
fraction for simple cubic, BCC, sheet gyroid, sheet I-WP and a gyroid--primitive hybrid, from the
skill's homogenizer. (b) The three best cells returned by search for $E_1=E_2=0.06$, $E_3=0.15$;
anisotropy comes from two periods along $x$ and $y$.}
\label{fig:lattice}
\end{figure}

The homogenizer reproduces a solid cell exactly and matches the exact $C_{33}$, $C_{44}$ and
$C_{11}$ of two-phase laminates to a relative error below $10^{-6}$ across several modulus
contrasts, Poisson ratios and phase fractions. A sheet gyroid at volume fraction 0.25 gives
$E/E_s=0.086$ with exact cubic symmetry. Figure~\ref{fig:lattice}(a) compares stiffness with
simple-cubic and BCC strut lattices: at equal density the TPMS and hybrid sheets are about
$5\times$ stiffer than BCC and $3\times$ stiffer than simple cubic in its weakest direction, though
simple cubic is $1.3\times$ stiffer along its own axes---which is why the skill reports the
directional minimum.

For one-off designs the skill searches a homogenized dataset and refines sheet thickness and
level-set bias. For the anisotropic target $E_1=E_2=0.06$, $E_3=0.15$ (relative to the solid) the
search and refinement took two minutes on a laptop CPU, excluding the one-time creation of a
300-cell dataset, and returned a gyroid--diamond--double-diamond hybrid with
$E=(0.057,\,0.058,\,0.151)$ at volume fraction 0.26 (Fig.~\ref{fig:lattice}b).
Table~\ref{tab:lattice-res} shows the sensitivity of the rounded published specification to voxel
resolution; the largest axial-modulus change from $32^3$ to $48^3$ is 2.2\%.

\begin{table}[t]
\centering\small
\caption{Resolution sensitivity of the lattice found by search (rounded specification: H3
gyroid--diamond--double-diamond, $t=0.744$, $h=0.209$, $a=-0.187$, periods $(2,2,1)$).}
\label{tab:lattice-res}
\begin{tabular}{rrrrr}
\toprule
Voxels & Volume fraction & $E_1/E_s$ & $E_2/E_s$ & $E_3/E_s$\\
\midrule
$24^3$ & 0.2549 & 0.0551 & 0.0556 & 0.1406\\
$32^3$ & 0.2605 & 0.0578 & 0.0573 & 0.1473\\
$48^3$ & 0.2598 & 0.0573 & 0.0577 & 0.1506\\
\bottomrule
\end{tabular}
\end{table}

The generative route adapts the paper's training recipe at reduced scale. A first run on a small
dataset (300 cells, $16^3$, CPU) produced unstructured fields; four changes produced structured
samples: training on the periodic grid with circular padding (which makes every generated cell
exactly periodic), a sinusoidal-plus-MLP time embedding in place of a per-step lookup table,
scaling the conditioning vector inside the encoder, and a longer learning-rate plateau patience.
Stiffness error after re-homogenization remained 200--280\% at this scale, against 3723 cells at
$64^3$ on four GPUs in the paper. The skill therefore uses search as its default design route and
accepts generated cells only after re-homogenization.

\subsection{Printed rockets (RocketSmith)}
\begin{figure}[t]
\centering
\includegraphics[width=0.62\textwidth]{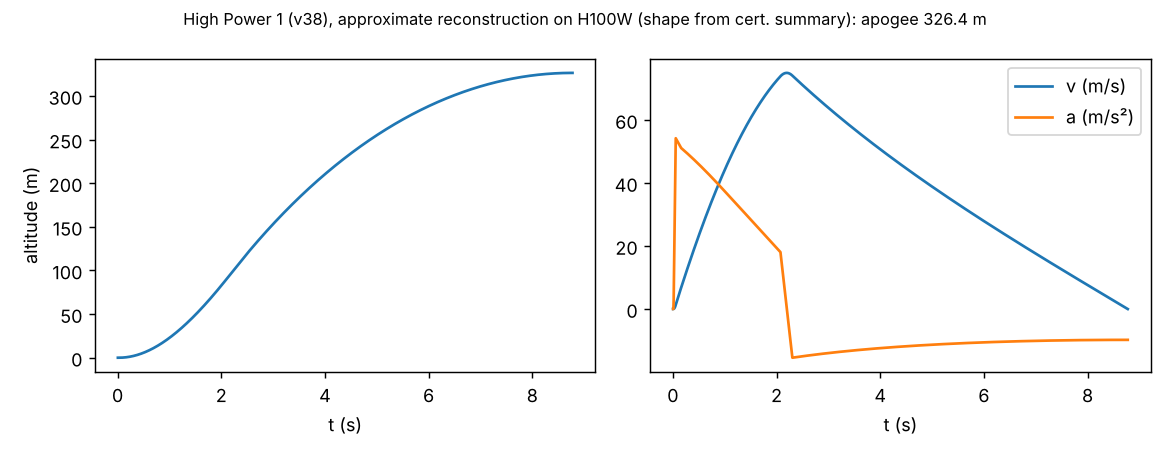}
\caption{\skill{rocketsmith}: simulated flight of the reconstructed HP1 rocket from the skill's
Barrowman stability model and one-dimensional flight simulator; apogee differs by 3.7\% from the
stored OpenRocket reference.}
\label{fig:rocket}
\end{figure}

RocketSmith~\cite{pak2026rocketsmith} designed four FDM-printed Class~2 rockets that all flew
stably, with altimeter apogees at 80--84\% of predictions. The skill couples a Barrowman stability
calculation~\cite{barrowman1967} to a one-dimensional flight simulator and ships approximate
reconstructions of two of the rockets. With $C_D=0.45$ it predicts 326.4~m and 75.0~m/s for HP1
against the stored OpenRocket~\cite{niskanen2009openrocket} references of 338.8~m and 76.4~m/s, and
567.6~m and 113.7~m/s for HP4 against 570.1~m and 115.2~m/s---apogee differences of 3.7\% and 0.4\%
(Fig.~\ref{fig:rocket}). Parameters the paper does not report were set or sized by the skill: HP1's
fin span (162~mm) was sized to the paper's liftoff stability of 1.26 calibers, so agreement on
stability is by construction, while the flight comparison is the untuned check.

Fitting an effective drag coefficient to the measured apogees of 276~m and 479~m gives
$C_D=0.84$ and $0.76$, well above the 0.4--0.5 typical of smooth commercial airframes. This
calibration describes those flights; surface finish, geometry differences, mass uncertainty and
flight conditions are possible contributors, and the skill exposes calibration as an explicit step
so that the fitted value can be carried to the next flight and tested there.

\subsection{Porosity from in-situ thermal images (ThermoPore)}
\begin{figure}[t]
\centering
\includegraphics[width=\textwidth]{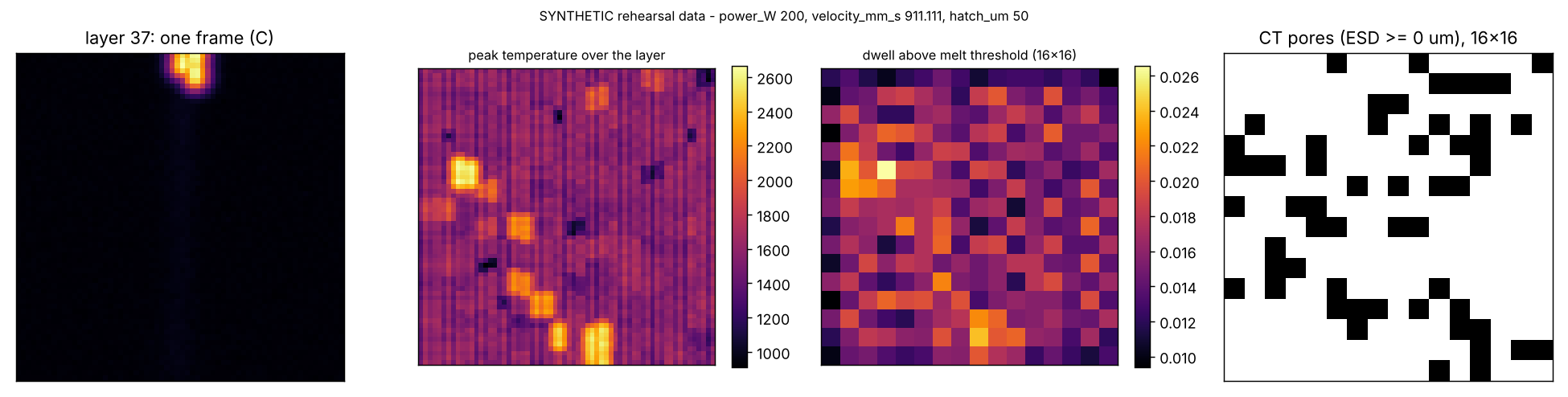}
\caption{\skill{thermal-porosity}: one layer of a synthetic rehearsal build. From left: a single
pyrometry frame with the melt pool mid-raster; peak temperature over the layer; dwell time above the
melt threshold pooled to the $16\times16$ label grid; and the synthetic pore labels used as ground
truth.}
\label{fig:thermo}
\end{figure}

Qualifying a laser powder-bed fusion part usually waits for X-ray computed tomography (CT).
ThermoPore~\cite{pak2024thermopore} asks how much of that answer is already contained in the
pyrometry recorded during the build~\cite{everton2016review}. For each 30~$\mu$m layer it pairs 200
thermal frames with the CT pores in that layer and trains a CNN that counts pores ($R^2=0.57$,
RMSE 7.84 in the source paper) and a video vision transformer~\cite{dosovitskiy2021vit,arnab2021vivit}
that localizes them on a $16\times16$ grid (mean IoU 0.32). The skill packages the workflow around
the models, which is where most practical effort goes: empty-frame filtering, resampling each layer
to 200 frames, alignment of CT and thermal coordinates with an explicit check image,
equivalent-diameter thresholds for which pores count, and splits within each process step as in the
paper (Fig.~\ref{fig:thermo}). It adds interpretable tree-based baselines on per-cell peak
temperature, dwell time above the melt threshold and thermal roughness, as a reference that learned
models must beat.

Because the paper's pyrometry and CT data are not public, the skill ships a synthetic rehearsal
build with the same structure (10 process steps, 16 layers each, 200 frames per layer; 120 training
and 40 test layers). On it the baseline reaches counting $R^2=0.44$ (RMSE 10.1) and localization IoU
0.27. These outputs carry a synthetic-data flag and establish pipeline behaviour, not accuracy on a
physical build. With a user's own build, the agent can simply be asked which layers and regions
should be sent to CT first.

\subsection{Sequential process optimization (SLS)}
For polymer selective laser sintering~\cite{pak2026sls}, the skill keeps a build log of 21
published batches across three materials, reduces ASTM D638/D790 tensile and flexural tests, and
suggests the next batch with a bounded Gaussian-process expected-improvement step~\cite{jones1998ego}.
This numerical suggester is an adaptation of the source agent workflow: it excludes results flagged
as non-reproducible, keeps the surface temperature at least 3~$^\circ$C below the melt peak and
limits how far one batch can move from the best so far.

To test it without hindsight we replayed the PA12 GF campaign, removing all later batches before
each of 11 decisions (A--C through A--M) and asking for three ranked suggestions with strength as the
objective. None of the 33 suggestions matched the next batch on all three settings (fill energy,
overlap and surface temperature). Agreement was directional: after A--C every suggestion moved to
50\% overlap, as batch D did, and after A--E the suggestions included 28~mJ/mm, the fill energy
later used in batch I. Because suggested batches that were never printed have no outcome, the
replay cannot show that following the suggester would have performed better; its role is to propose
a bounded next step for the engineer to judge by the printed result.

\subsection{Equation discovery (LLM-SR)}
LLM-SR~\cite{shojaee2025llmsr} lets a model propose equation skeletons as code while an optimizer
fits the constants. The skill evaluates proposals in a time-limited, isolated Python subprocess
after static source checks (a guard against mistakes, not a security sandbox) and fits constants by
BFGS. On the LLM-SR stress--strain benchmark, the eight-parameter skeleton
\begin{align}
\hat\sigma(\epsilon,T)=(p_0+p_1T+p_6T^2)
\left[1-\exp\{-p_2(1+p_7T)\epsilon\}\right]-(p_3+p_4T)|\epsilon|^{p_5}
\end{align}
reaches normalized MSE 0.021, 0.025 and 0.074 on the training, in-domain and held-out-temperature
splits (2161, 1442 and 738 rows), against 0.161, 0.150 and 0.680 for a cubic polynomial and 0.380,
0.377 and 1.31 for linear regression (Table~\ref{tab:eq-replay}); the original repository's evaluator
gives the same training error. Every term has a physical reading: an elastic rise, a saturation
stress that falls with temperature, and necking softening. The held-out temperature (0.667) lies
inside the training range (0.067--1.0), so the result shows generalization to an unseen condition
by interpolation, not extrapolation.

The replay also exposed a methodological flaw that we corrected in the skill: its prompt builder
had shown the held-out score to the model while it proposed equations, so that score could not
serve as an untouched final test. The skill now uses the in-domain split for feedback and
selection, hides the held-out split from every search command, and reveals it once, through an
explicit \texttt{final} step, for the equation the user has frozen; tests enforce this behaviour.

\begin{table}[t]
\centering\small
\caption{Fixed-skeleton replay on the LLM-SR stress--strain benchmark (normalized MSE). The scores
reproduce the numerical fit of the discovered skeleton, not the original search campaign.}
\label{tab:eq-replay}
\begin{tabular}{lrrr}
\toprule
Model & Train & In-domain & Held-out temperature\\
\midrule
Discovered eight-parameter skeleton & 0.02096 & 0.02464 & 0.07362\\
Cubic polynomial & 0.16141 & 0.14952 & 0.67952\\
Linear regression & 0.38018 & 0.37661 & 1.31432\\
\bottomrule
\end{tabular}
\end{table}

\subsection{Independent rerun and evidence scope}
\label{sec:rerun}
All case-study numbers were recomputed in a separate rerun from library commit \texttt{46cfac2} and
LLM-SR commit \texttt{41c2123} (macOS 15.7.7, arm64; Python 3.9.6, NumPy 2.0.2, SciPy 1.13.1,
scikit-learn 1.6.1, PyAMG 5.3.0), by direct script execution without modifying the implementations.
The five test suites passed all 68 assertions that can run without optional dependencies (15
rocket, 17 lattice, 12 thermal-porosity, 12 SLS and 12 equation-discovery); CadQuery export and
PyTorch checks were skipped rather than counted. Table~\ref{tab:rerun-scope} summarizes what each case
establishes and where its evidence stops. The rerun exercises the tools; it does not measure skill
routing, natural-language reliability or voice operation, which are the subject of the controlled
evaluation proposed below.

\begin{table}[t]
\centering\small
\caption{Evidence scope of the five case studies.}
\label{tab:rerun-scope}
\begin{tabularx}{\textwidth}{@{}>{\raggedright\arraybackslash}p{2.2cm} >{\raggedright\arraybackslash}X >{\raggedright\arraybackslash}X@{}}
\toprule
Workflow & What the evidence establishes & Where it stops\\
\midrule
Lattice & Exact solid and laminate agreement; resolution sensitivity; search result re-homogenized & No full-scale diffusion training or physical coupon tests\\[3pt]
RocketSmith & Apogee and velocity within 4\% of stored OpenRocket references & Reconstructed geometry; OpenRocket not rerun; no new flight\\[3pt]
ThermoPore & Complete preprocessing, labelling and baseline pipeline on synthetic data & No source pyrometry/CT data; deep-model accuracy not reproduced\\[3pt]
SLS & ASTM data reduction; 11 chronological replays with directional agreement & Retrospective; no outcomes for unprinted suggestions\\[3pt]
LLM-SR & Discovered skeleton and baselines refitted on the original benchmark files & Original model-driven search not rerun; held-out split is interpolation\\
\bottomrule
\end{tabularx}
\end{table}

\section{Limitations and future work}
\label{sec:limits}
The case studies verify selected executable components and document how each research method was
adapted; they do not yet show library-wide agent reliability or an improvement over an agent with
the same tools but no skills, and they do not reproduce all 19 source papers. Synthetic rehearsals
test pipeline behaviour and cannot support accuracy claims on physical builds, and learned
components rehearsed at small scale remain unvalidated at the source papers' scale.

The next step is a controlled evaluation. Task definitions and acceptance criteria will be frozen
before execution; the same agent with the same tool access will be run with and without skills
under equal budgets and repeated across seeds; and design, analysis and manufacturing tasks will be
scored on constraint satisfaction, numerical correctness, failed or skipped checks, unsupported
claims, completion time and cost, including spoken requests. An ablation that removes the
verification instructions will measure their contribution.

Several skills depend on external software (OpenFOAM, FreeCAD, Thermo-Calc), licensed data or
hardware (a Klipper printer, a powder-bed machine); they ship validators, mock devices or rehearsal
data, but real use requires the tools. Instructions make assumptions and checks visible, but
whether an agent follows them, and whether the checks are adequate, must be measured. Skills
inherit the limits of their methods and do not replace an engineer's sign-off on safety-critical
decisions.

\section{Conclusion}
Mechanical Engineering Skills packages 33 engineering capabilities, including 19 adaptations of our
research, as instructions, executable tools and verification procedures that a general agent can
discover, load and use. The central idea is simple: let the model plan and explain, let tested code
compute, and return every result with its check. Because the interface is natural language, this
breadth is available in one conversation, by text or voice, without learning each tool, training
models or running them locally. Writing the skills and rerunning their case studies also made each
method's evidence explicit---and exposed and fixed a flaw in one evaluation---which suggests that
a skill is a useful standard for research reproducibility as well as a delivery format. The library
is open-source under the MIT license; we invite other groups to contribute skills and to help
validate them.

\paragraph{Availability.}\sloppy\ Code, tests and tutorials:\\ \url{https://github.com/BaratiLab/mechanical-engineering-skills};\\ website: \url{https://baratilab.github.io/mechanical-engineering-skills/}.\\ Install all skills with
\texttt{npx skills add BaratiLab/mechanical-engineering-skills --all}.

\bibliographystyle{unsrtnat}
\bibliography{refs}

\end{document}